\documentclass[10pt,a4paper]{article}
\pdfoutput=1
\usepackage[hidelinks]{hyperref}
\usepackage{cite}
\usepackage{doi}
\usepackage[margin=1.5cm]{geometry}
\usepackage[utf8]{inputenc}
\usepackage[T1]{fontenc}
\usepackage{graphicx}
\usepackage{multicol}
\usepackage{changepage}
\usepackage{caption}
\usepackage{textcomp}

\begin{document}

\begin{center}

\textbf{\Large{Synaptic clock as a neural substrate of consciousness}}
\vspace{0.5cm}

Bartosz Jura \\
\vspace{0.1cm}
\textit{Department of Cognitive Neuroscience, Institute of Applied Psychology, \\
Jagiellonian University, Cracow, Poland} \\
\vspace{0.1cm}
Correspondence: bartosz.jura@uj.edu.pl
\vspace{0.5cm}

\begin{adjustwidth}{2cm}{2cm}
	In this theoretical work the temporal aspect of consciousness is analyzed. We start from the notion that while conscious experience seems to change constantly, yet for any of its contents to be consciously perceived they must last for some non-zero duration of time, which appears to constitute certain conflict. We posit that, in terms of phenomenological analysis of consciousness, the temporal aspect, and this apparent conflict in particular, might be the most basic property, likely inherent to any conceivable form of consciousness. It is then outlined how taking this perspective offers a concrete way of relating the properties of consciousness directly to the neural plasticity mechanisms of learning and memory, and specifying how exactly subjective experience might be related to processes of information integration. In particular, we propose synaptic clock to constitute a content-specific neural substrate of consciousness, explaining how it would correspond to this temporal aspect. Then, we propose a viewpoint, in which moments of subjective time have different durations, depending on the type of information processed, proportional to the time units of corresponding synaptic clocks, and being in principle different for different brain regions and nervous systems in different animal species. Relation and possible contributions of this viewpoint to the extensional model of time consciousness are discussed. Finally, we consider the two alternative views on the structure of consciousness, namely a static and a dynamic one, and argue in favor of the latter, proposing that consciousness can be best understood if change is considered its only dimension. \\
\vspace{0.2cm}
\end{adjustwidth}
\vspace{0.5cm}

\end{center}

\begin{multicols}{2}

\section{Introduction}

	It has been long observed that an association between two percepts, thoughts, or contents of conscious experience in general, can be created in a form of relational memory if they occur in temporal proximity. It is unclear, however, what exactly does it mean for two such events in one's mental life to occur in temporal proximity. How long such a temporal distance between two events of the same or different modalities can be, or, first of all, how can the time of the flow of conscious experience be measured? Is it adequate for it to be expressed and measured in the units of physical time, i.e., the ones in which time is measured by conventional, physical clocks (like those used to time neural activity in standard brain imaging experiments)? \par
	In order to address these questions in a strict manner, it is necessary to determine what exactly is the relation between subjective experience, i.e., consciousness, and, on the other hand, objective reality, part of which the physical clocks are. In other words, assuming that one's subjective experience is always related to some physical processes, to have a definition of what exactly constitutes a physical substrate of consciousness, what within that substrate constitutes a given experience with some specific contents, and what kind of rule, if any, governs the progression and succession of various contents. A notion, common especially outside the field of 'temporal consciousness’ research \cite{Dainton2018,Kent2021}, that subjective time indeed can be expressed in the units of what is considered physical time, seems to be based on an assumption that this relation is of a one-to-one nature, namely that any one 'tick' of some underlying physical system will always correspond to one 'tick' of subjective experience. However, as so far there seems to be no evidence for this to be the case (and it is hard to imagine how this sort of dependency could ever be verified experimentally), a reasonable approach to take is to assume that this relation is not necessarily of such a straightforward nature. \par
	Here we take this approach and argue that, in case of biological systems, combining certain observations coming from the two domains, specifically the fact of linking of subjectively experienced events that occurred in temporal proximity \cite{Howard2015}, with the knowledge about putative neural mechanisms of learning and memory and, in particular, of linking memories over time \cite{Sajikumar2004,Cai2016}, might be informative in terms of elucidating what are the content-specific neural substrates of consciousness. \par
	Conscious experience is usually assumed to be confined to momentary presents–a moments of ‘now-ness’. This notion, however, poses serious difficulties for explaining how any perception of change (e.g., a melody), or a lack of change (e.g., a silence), is possible, with different models trying to reconcile this two seemingly contradictory properties being attributed to consciousness \cite{Dainton2018}. Some authors even argue that such perceptions of change might actually be illusions \cite{Dennett2018}. \par
	Here, we analyze the temporal aspect of consciousness, and consider an idea that a content-specific neural substrate of consciousness can be found directly in the mechanisms of neural plasticity underlying the processes of learning and memory. In particular, we put forward arguments showing why synaptic clock, a hypothetical mechanism based on activity-dependent synaptic plasticity we recently proposed \cite{Jura2019}, could constitute such a substrate. This proposal is analyzed in light of certain aspects of different current theories, in particular theories relating consciousness to processes of learning and memory and plasticity of the nervous system \cite{Flohr1991,Cleeremans2011,Mogensen2011,Lamme2018}, as well as theories that relate consciousness to processes of activity (or, information) integration, like the information integration theory (IIT) \cite{Tononi2004,Koch2016}. IIT starts from a set of axioms derived from a phenomenological analysis of consciousness, and posits that consciousness is related to the capacity of a system for information integration. We start with a notion that consciousness seems to be inherently temporal in nature, that is notion laying somewhat beyond the scope of attention of IIT and many other theories (see also \cite{Kent2021} and for an overview of theories \cite{Northoff2020}), and argue that, nonetheless, our approach might be particularly informative for attempts of defining how exactly consciousness could be related to processes of information integration. \par
	In the last section, we consider consciousness in more general terms, showing how analysis of the actual meaning of this 'temporality' of consciousness can be informative for attempts of defining its 'place' within the physical world, or the exact relation between one's subjective experience and the objective reality. In particular, this approach allows to propose a specific interpretation of a view according to which the brain actually learns to be conscious \cite{Cleeremans2011}. \par

\section{Synaptic clock as a neural substrate of consciousness} \label{section1}

\subsection{Temporal aspect of consciousness considered from an evolutionary-ecological perspective}

	From the perspective of individual organisms, the specific sequences of events over their lifetime are unique for organisms from different species as well as for every individual within a given species (Figure \ref{Fig1}A, top part). However, all those events always take place in time, they take time, and the distance between consecutive events can always be considered as temporal in nature. They can be marked, in a sense, on lines, that would look identical for every organism (Figure \ref{Fig1}A, bottom part). To traverse any spatial distance, or learn a space, it always takes some amount of time, whereas time passes even without one changing its spatial position. \par
	From this perspective, interactions of an organism with its surroundings can be depicted as a loop (Figure \ref{Fig1}B), in which sensory cues that the organism receives at any given moment of time depend on, that is, are feedbacks for, the motor actions it performed in previous moments of time. It should be noted that by "motor actions" we mean here also a lack of any actions, that is, for instance, a decision made by the animal, caused by it having spotted a predator, to stand still and not make any movements. And by "feedbacks", analogously, we mean here also a lack of any changes in the outside world, that is, for instance, the predator not moving towards the animal after it did not make any rapid movements that would attract the predator’s attention. Defined in this way, the \begin{center}
\fbox{\includegraphics[width=\linewidth,keepaspectratio]{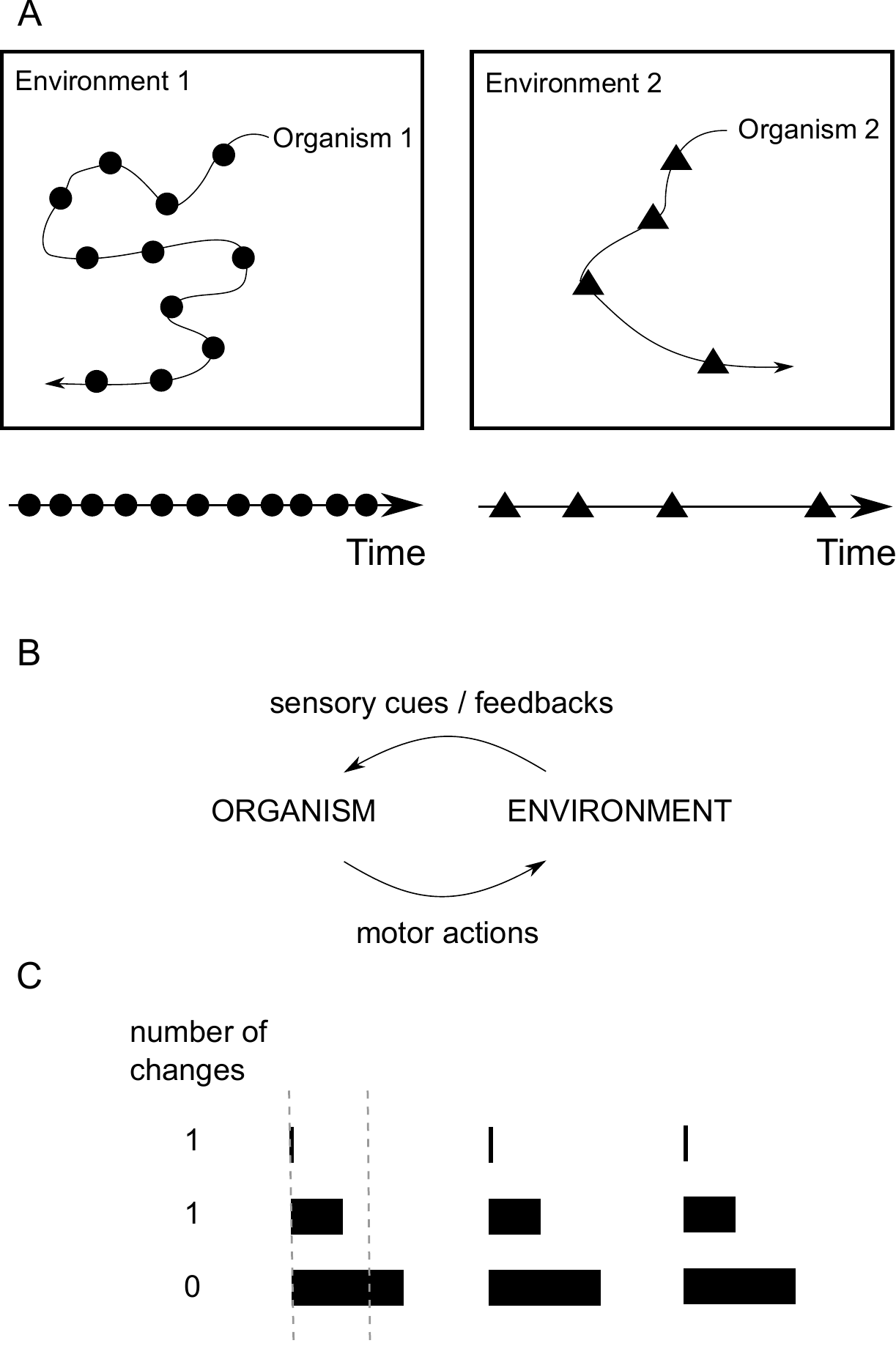}}
\captionof{figure}{Temporal aspect of consciousness considered from an evolutionary-ecological perspective. \textbf{(A)} Time as an universal dimension in interactions of organisms with their respective environments. \textbf{(B)} Organism--outside world feedback loop of interactions. \textbf{(C)} Comparison of clock models with discrete vs. persistent ticks. Judgments of the duration of a time interval, according to different clocks, all ticking at the same rate--a classic one with discrete ticks (top), and two with persistent ticks of different non-zero durations (middle and bottom), as in the synaptic clock model. Although the total number of ticks generated over the interval as a whole is equal according to each of the clocks (as would be assessed retrospectively), what differentiates these clocks is the timing of the short sub-interval (between gray dashed lines), with different numbers of changes occurring, which is due to different durations of the persistence of their respective ticks. \label{Fig1}}
\end{center}motor actions are always associated with some underlying neural activity (be it even such resulting in the suppression of execution of some specific movement, or continuation of a previous one), as well as some sensory cues or feedbacks will be at all times processed and evaluated and will be related to some neural activity, in order to assess what outcomes the motor actions had brought for the organism (resulting in, at least, fluctuations of what could be considered a baseline-level emotional state). \par
	In this sense time can be therefore considered a universal ecological dimension, offering an evolutionary-ecological perspective suggesting that some universal rules might have developed utilizing this fact, as all aspects of organisms’ functioning are, in a sense, subordinated to this dimension. \par
	The perspective outlined is in particular the perspective of an individual's subjective experience. Namely, along a line like the ones on Figure \ref{Fig1}A proceed one's mental states, i.e., conscious experiences, occurring, in a sense, in between the generation of motor actions and detection of sensory cues/feedbacks, as is also the case for neural activity with which they would be related. What an individual experiences at any given moment is the flow of contents of experience, or, in other words, the flow of subjective time. \par

\subsection{Experience of time flow as the most basic aspect of consciousness}

	It seems to us that reasoning like the above can be extended even further, and it can be posited that the most basic aspect common to the conscious experience of any system, if only it is endowed with consciousness, is the very fact of the perception of time flow, as everything else might be, conceivably, perceived differently. Namely, for instance, perception of colors: what I perceive as color red can be completely different from what you perceive as color red \cite{Byrne2020}. We use the same word to describe it--"red"--and we agree that what we both are seeing is "red", but what each of us actually sees, and thus means by "red", can be completely different (e.g., your "red" could be my "green", and \textit{vice versa}) and there is no way to communicate it. The same applies to all other elements of mental life in general. Moreover, in terms of different percepts in a given individual, what we can only say about them is, for instance, that what I perceive as color red is \textit{different} than what I perceive as color blue, or that color red is different than a percept of a sound, but I am unable to say \textit{how} exactly they are different, or what exactly makes them so. Perhaps, if we assume that the ability to have those percepts is a product of evolution, the corresponding qualities, as far as such correspondence can be defined, will be similar in individuals within the same species (although, in fact, there seems to be no utilitarian reason whatsoever for this to be the case), but they might likely be very different in individuals from distant species, e.g., in humans and in flies, respectively. This reasoning leads eventually to the conclusion that the most basic aspect of conscious experience that will be common to any conceivable form of consciousness, in different systems, is the very feeling of the fact of time flow, i.e., constant flow of conscious experience. \par
	Assuming that consciousness is subject to evolution and might be present in simpler forms also in simpler systems, then even in the case of a simplest conscious system conceivable, capable of having only an experience of "this rather than not this" \cite{Tononi2015} (e.g., "light rather than not light"), it will experience the flow of time. That is, at least, an experience of "this rather than not this" actually appearing and/or disappearing. We assume that such an experience of time flow is real (rather than a mere illusion) \cite{Wittmann2009,Montemayor2021}. \par
	However, despite this constant flow (which cannot be voluntarily stopped--see the Discussion for an extended argumentation pertaining to this point), a percept, thought, or any content of experience in general needs to last for some non-zero duration of time in a seemingly unchanged form in order to be consciously perceived (i.e., 'registered' in consciousness), which appears to constitute certain conflict. How to reconcile these two notions? How long exactly such contents persist in consciousness? \par
	We argue that such persistence is due to a form of immediate memory, analogously to the transient persistence of visual stimuli, that would persist in, or actually \textit{as} consciousness, for a certain amount of time, before it is 'replaced' by another content (and that can potentially proceed to be stored as a longer-lasting memory). Based on studies in the visual system, it is estimated that in humans such percepts last usually not less than tens of milliseconds and not more than hundreds of milliseconds \cite{Motanis2018,Tononi2015}. In other species this is assessed by indirect measures, and represented by the values of Critical Frequency of Flicker Fusion (CFF), determining the durations of the persistence of visual stimuli that leads to fusion of consecutive stimuli, which lay in the range from less than ten milliseconds (in insects), to below one hundred milliseconds (in turtles) \cite{Healy2013}. However, what about other types of mental states, beyond primary sensory perception, e.g., more abstract thoughts? How long they can, or should, last in different conscious systems that can possibly be thought of? The values of CFF display a specific species-varied pattern that can be attributed to certain evolutionary-ecological processes \cite{Healy2013}, suggesting that the persistence of visual percepts is precisely tuned in particular species. It thus seems reasonable to assume that the persistence of other types of contents of conscious experience will also be shaped evolutionarily, in accord with the visual system's CFF, so that they last as long as it is suitable for individuals from species living in given circumstances. That is, not too long and not too short, likely being proportional to intervals separating behaviorally relevant events in which a corresponding type of information is processed. \par
	That this is indeed the case is suggested by the variable speed of our subjective perception of time flow. The fact that intervals of clock time of a given objective duration appear to us as having variable duration may imply, that the contents of experience can persist for different amounts of time, making us unable to distinguish shorter fragments (sub-intervals) within intervals in which those contents were fixed (as illustrated on Figure \ref{Fig1}C). This phenomenon seems to depend on the type of information processed. For example, it seems that processing of more abstract information, like a spatial information about environment, related with the feeling of knowing "where I am at the moment", or when solving a mathematical problem, is related to contents of experience that persist for longer time periods, increasing the perceived speed of time flow (when it is assessed with reference to some indications of the objective clock time). While in principle a retrospective judgment of interval duration can be impacted by various memory distortions or ‘failures’, preventing an information about events that just occurred to be retrieved from a memory store, explaining thus variability of such a judgments \cite{Dennett1992,Howard2015}, we argue that the persistence of contents of experience continuously affects the perception of time flow, and the issue of its duration is worth addressing. \par
	The concept of synaptic clock, which in the cognitive domain is based on generalization of a rule based on the visual CFF to other aspects of brain function, describes how the persistence of different types of information could span broader time ranges, resulting in different 'CFF's, being subject to specific selective pressures \cite{Jura2019} (Figure \ref{Fig1}C). This thus situates the synaptic clock as a potential content-specific neural substrate of consciousness, that will correlate with the persistence of various contents in conscious experience, a hypothesis which we shall consider next. \par

\subsection{Synaptic plasticity as content-specific substrate of consciousness}

	Usually, when potential neural correlates of consciousness are being considered, including their content-specific subsets, what is thought of most naturally as a candidate mechanism is, essentially, neurons firing action potentials. Such correlates are sought primarily in patterns of neuronal firing, global or localized to specific systems, reflected directly by spikes or by oscillatory activities of field potentials, interpreted as effects of synchronized firing of groups of cells \cite{Koch2016,Northoff2020}. As neuronal firing is necessary for the nervous system and organism as a whole to function properly, it is perhaps also necessary for conscious experience as such, and thus its contents, to be possible to occur in such organisms. It will always be there, in one way or another correlating with experience. However, there seems to be no evidence showing why neuronal firing as such should be in a privileged position in this regard, and should constitute a substrate directly related to some specific contents of a conscious experience. Instead, there are some features of synaptic activity, and of synaptic plasticity as represented in the synaptic clock hypothesis in particular, that we will now discuss, and that allow to see it as more well-suited (and having more to 'offer', in terms of a broader repertoire of possible mechanisms in play) as a candidate for a content-specific neural substrate of consciousness, than neuronal firing \textit{per se}. \par
	Synaptic clock hypothesis assumes a brain-wide distribution of default durations of persistence of transient, activity-dependent synaptic traces. By synaptic trace it defines an event of transient synaptic plasticity, based on a generalized notion of synaptic tagging, constituting a memory trace of previous synaptic input, or, more generally, of previous synaptic activity. As an instance, we will consider here a general case of an event of synaptic tagging (as is studied mostly in rodent hippocampal cells), triggered by weak synaptic stimulation, mediated by activation of NMDA receptors, associated with protein synthesis-independent early phase of Long-Term Potentiation (LTP), or Long-Term Depression (LTD), and which can be transformed into a late phase of LTP (l-LTP) or LTD, with the late phase being dependent on (1) plasticity-related proteins (PRPs), needed to actually implement a longer-lasting synaptic change, and (2) some external (i.e., extracellular), more global reinforcing signals (e.g., dopamine; which stimulate/modulate the production of PRPs and/or directly modulate the synaptic change), that both can be delivered to the synapse with a delay after an initial synaptic event \cite{Sajikumar2004}. \par
	The arguments (approaching the issue from different angles) could be as follows:
\begin{itemize}
	\item Consciousness is informative, i.e., constituted always by some specific content. And it can be stated, in general terms, that whereas the role of neuronal firing and synaptic transmission is to transmit signals, the role of synaptic plasticity is to encode and store information within a network. It is then reasonable to assume that information that is actually being encoded by particular synapses will correspond to the contents of an ongoing conscious experience;
	\item Every one content of conscious experience persists in time, for some non-zero duration of time. It can be stated, in general terms again, that whereas an action potential is a discrete event, meaning that it does not persist anywhere but propagates from one place to another, synaptic memory trace, in contrast, does persist for a prolonged period of time in a well-localized site. It is a persistence of a previous activity state;
	\item Conscious experience changes constantly. And while synaptic trace is something that persists, it also is, by definition, a synaptic \textit{change}. Together with the point above it thus directly corresponds to our main issue as discussed above (and elaborated on below, in the last section);
	\item NMDA receptors, as mediating an activity-dependent formation of dynamic neural assemblies, have already been proposed to be implicated in the occurrence of phenomenal states \cite{Flohr1991};
	\item According to IIT, consciousness is related to information integration. And synaptic trace, in itself, integrates information. Specifically, on the one hand, taking the example of spike-timing-dependent Hebbian plasticity, as embodied in the process of LTP mediated by NMDA receptor activation, it is triggered by, and necessitates, the integration of a specific pattern of presynaptic activity with postsynaptic depolarization, all with specific timing, which altogether might determine whether the synapse will undergo LTP or LTD, which in turn determines its effect on subsequent network activity, routes of signal transduction, and eventually behavior. Besides that--and, for our perspective, more importantly--the late phase of plasticity (l-LTP) is dependent on specific PRPs and neuromodulatory signals, delivery and action of which depends on the behavior of animal and patterns of neural activity in a prolonged time window after, as well as before, the initial synaptic stimulation. The final fate of synaptic change, and future activity of the circuit it is a part of, is thus determined by, i.e., integrates, a combination of numerous events and specific patterns of activity;
	\item Relating to the above point, a substrate of consciousness is sought, among other mechanisms, in recurrent interactions, that might be involved in certain top-down signaling processes in the brain \cite{Lamme2018}. As this type of activity is especially likely to activate NMDA receptors, whose activation at the moment of a recurrent signal would be enabled due to the cell's membrane depolarization caused by a preceding stimulation by feedforward signal, and NMDA receptors are implicated in synaptic plasticity, like LTP, it is thus hypothesized that consciousness might be not due to recurrent interactions \textit{per se}, but rather due to a sequence of events that they trigger, starting with the initiation of synaptic plasticity \cite{Flohr1991,Lamme2018}. This would implicate the proposed mechanism of synaptic clock into effects associated with recurrent interactions in the brain;
	\item IIT posits that even inactive, i.e., non-firing, cells can contribute and shape the contents of a conscious experience (but not cells that are artificially, e.g., pharmacologically, blocked--i.e., functionally detached from the rest of a network) \cite{Tononi2015}. In line with this notion, synaptic traces are assumed to persist as well in neurons which are silent, in terms of not generating action potentials at the moment. This could limit a possible pool of inactive cells that should be taken under consideration, defining when a given cell can directly, i.e., positively, affect a conscious experience. That is, only when a synaptic clock that has been activated in that cell is still active. It seems more likely that only such a cells will directly affect an ongoing experience, as opposed to ones in which there is nothing that could be considered a trace, or 'evidence', of any activity that would make it involved in network activities and make it directly contribute to a conscious experience (on how to understand this ‘direct contribution’ we elaborate further below);
	\item Neural substrate of consciousness may not be fixed, restricted to some specific circuits or populations of cells, or specific patterns of neuronal activity, but rather be related to the performance of higher-level functions and as such be more 'flexible'--a view suggested by the plastic nature of nervous system, manifesting itself especially in the reorganization of circuitry underlying certain cognitive or behavioral functions after brain injury \cite{Kupers2011,Mogensen2011}. In line with this, regardless of which exact areas or circuits will be involved in execution of a particular function and processing of relevant information, corresponding synaptic clocks might possibly be activated in those circuits;
	\item Consciousness has been hypothesized to be related to neural activity that underlies learning, as opposed to activity underlying events that are yet too 'weak' or 'unimportant' to require learning, or, on the other end, activity underlying performance of already-learned automatic functions that does not need to be updated \cite{Schrodinger1974,Cleeremans2011}. Since the process of learning itself, as well as memories that were created in some specific life circumstances and are then recalled in a given conscious experience, are always associated with some emotional state, or value, according to this view consciousness is thus tightly related to emotions \cite{Cleeremans2011}. And since emotions are related to neuromodulatory effects in the brain, with neuromodulators, like dopamine, acting on synaptic traces, possibly with a delay, and thus modulating learning–synaptic clock can be thus seen as a mechanism which directly receives, or 'senses', the information about an outcome that a preceding or ongoing neural activity has brought for the organism, as conveyed by the emotional state associated with a given conscious experience (a notion on which we shall elaborate in the last section).
\end{itemize} \par
	Based on these points, synaptic clock is thus a concrete candidate for a content-specific substrate of consciousness, as it (1) is capable of accounting for certain basic phenomenal properties of consciousness, related to its temporal nature, (2) relates consciousness to specific neural mechanisms of learning and memory (which puts it in line with certain aspects of theories relating consciousness to processes of learning and memory and/or recurrent signaling), and (3) defines exactly (i) where, and (ii) in what (biological) senses, information will be integrated strongly. \par
	To specify, what will constitute a content-specific substrate of consciousness is in this view not a single one but rather all currently active synaptic clocks, with the focus of attention (externally or internally directed) affecting which circuits have greater contribution to the actual contents of an experience, with the ‘contribution’ of specific circuits, cells, or synapses, to be determined, we argue, based on whether and how they affect the overall ongoing experience of time flow. In the following section we will further specify how this should be understood from a broader perspective of a network, part of which the synapses are and activity of which the synaptic plasticity affects. \par

\section{On the duration of a single moment of subjective time} \label{section2}

\subsection{Subjective moments of time with different durations}

	As is inferred mostly from animal experiments, cellular and synaptic mechanisms that are responsible for 'allocation' and initial encoding of memory of particular events, determining which cells and synapses will be recruited to a particular memory trace (defined as a pattern of activity of specific neuronal ensemble), are also the ones that lead to linking memories of events that occurred close in time \cite{Rogerson2014,Cai2016,Sehgal2018}--which suggests that mechanisms of neural plasticity underlying learning and memory will in themselves lead to creation of associations, in a form of relational memory, between contents of experience that occur in temporal proximity. However, then the questions are, (1) how to measure and express this temporal distance for contents of conscious experience of the same or different modalities, and what is its maximum value to which creation of such associations is restricted, and (2) what it can tell us about the exact nature of content-specific substrates of consciousness? We propose that a pragmatic approach to these questions (which, however, should be seen as only a practical approximation) is to assume that an association between different contents of conscious experience can be formed only if they occur at the very same moment of subjective time, with moments of subjective time having different durations, as expressed in ‘neural’ time, depending on the type of information processed (Figure \ref{Fig2}). \begin{center}
\fbox{\includegraphics[width=\linewidth,keepaspectratio]{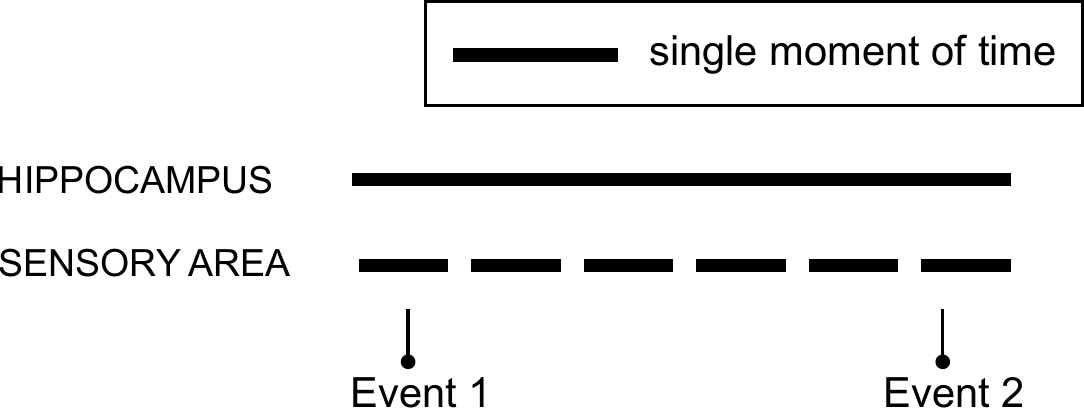}}
\captionof{figure}{Duration of a single moment of subjective time. Durations of single moments of subjective time, proportional to the time units of synaptic clocks in different brain regions, in which different types of information are processed, constituting temporal windows for the formation of associations between events. Two events, certain aspects of which are processed by both regions, being simultaneous as judged from the perspective of the hippocampus, are not simultaneous when judged from the perspective of the sensory area. Values presented are hypothetical, and not necessarily in scale. \label{Fig2}}
\end{center}\par
	It can be noted that, in general terms, the transmission of effective signals in the nervous system, and correspondingly the interactions of organisms with their environments, are directional, i.e., neural activity always leads eventually to some behavioral output (in the sense as defined above--be it even suppression of a motor action). Assuming that any one conscious experience is always associated with some neural activity (that is, in organisms endowed with both nervous system and consciousness), then the most basic overall situation of a biological system, being part of some environment, having some specific experience, can be described as: behaving in a certain way when it experiences "this rather than not this". In this view, the effect of any experience-dependent synaptic plasticity elicited in such situation, related with the contents of concurrent experience, will be to alter the possible patterns of neural activity and consequently, due to the directionality, behavioral output. From this perspective, instead of a notion of memory traces as faithful representations of particular events, one should expect memory to behave more like a fluid \cite{Hardt2020}, with neural plasticity serving primarily the ultimate goal of the organism, namely of adjusting its behavior to particular circumstances it finds itself in. In such elementary case, to create an association between different contents of experience, in a form of relational memory, primarily due to plasticity of pre-existing synaptic connections within an overlapping neural substrate, will be to make the patterns of neural activity and behavioral output related to those contents convergent, i.e., more similar to each other than they had been before the association was formed. This reasoning, as it seems, can be extended also to complex systems \cite{Miyashita1988}. \par
	As contents of conscious experience we include here all the content that constitutes a given experience at a given moment, no matter how 'weakly' one would be aware of its presence or contribution to the experience, or whether one is able to recall and reflect on that experience in a retrospective manner after some time has passed. That is, also contents constituted by processes like remembering events, or percepts, perceived in the past, no matter how distant, or cognitive manipulation of a given percept, as well as visualizing and thinking about future ones--all of which, in this view, constitute another experience in its own right, occurring at some present moment of time, and being a form of ‘immediate memory’, as defined above, related thus to some neuronal activity, synaptic stimulation and, consequently, formation and persistence of synaptic traces, i.e., activity of synaptic clocks. \par
	Subjective time flows with variable rate, as we know it from introspection, and it seems that it flows with yet different rates for individuals from different species \cite{Healy2013}. From the effective perspective of one's subjective experience, this can be attributed to an effect resembling the flicker fusion effect in visual system, with contents of experience persisting for some non-zero duration of time before they will disappear, making 'room' for other contents (as we argued already above). Such a persistence determines thus an irreducible unit of subjectively perceived time, not divisible into shorter fragments, constituting a single moment of subjective time, which, however, will be related to some portion of neural activity extended in time--a window of 'neural' time of non-zero duration. We posit that durations of such moments of subjective time will be proportional primarily to the time units of synaptic clocks, in principle different in different brain regions and across animal species, depending on the type of information processed by given synapses (determined, in general, by the anatomical and functional connectivity of a given region; approximate distribution of which was proposed in \cite{Jura2019}), and that those moments constitute temporal windows for the formation of associations between different events, aspects of which are processed by a given region. Synaptic tagging, being an early stage of activity-dependent synaptic plasticity, on which the concept of synaptic clock is based, allows for 'late-associativity'--a prolonged time window in which, presumably, associations can be created between events that are separated in time on the neural level \cite{Sajikumar2004}. Although, perhaps, the durations of those moments will depend also on other, complementary plasticity mechanisms involved in processing of successive events and memory 'allocation', like the transient cellular-level plasticity, i.e., altered neuronal excitability. Altogether, the discussed mechanisms will be causing certain 'inertia' after an initial neural activation, which will result in a ‘fusion’ (or ‘integration’) of the spatio-temporal neural representations of different events processed by a given region. As an instance, for hippocampus and a sensory area, as depicted in Figure \ref{Fig2}: a circuit in which the moments are longer, e.g., the hippocampus, will integrate information about the aspects of events that it processes for any two events that occur within its current moment of time, thus 'seen' by it as one event, even if one of those events occurred in a distant past according to the sensory area endowed with moments of shorter duration, i.e., with traces of activity left by that event in the sensory area having decayed long time ago, as judged from its perspective. \par

\subsection{How to infer the durations of subjective moments from the level of network activity}

	Subjective time understood as representing an order of events can be reconstructed from patterns of population- as well as single cell-level neuronal activity (as shown in particular in the lateral entorhinal cortex \cite{Tsao2018}). The present theory posits that if one extracted a (hypothetical) component of the evolution of a given activity pattern that could be attributed to plasticity of the neural substrate, then such components in different brain regions, or subpopulations of cells, would change with different rates, with synapses with shorter time units returning to putative baseline levels more rapidly after an initial stimulation event. \par
	One possible practical way of determining the putative durations of the subjective moments, would be thus to measure the similarity between network responses (a degree of overlap of the network activity patterns) to a pair of dissimilar stimuli applied at varied inter-stimulus intervals, and compare this dependence for stimuli of different modalities and corresponding brain regions (assuming that a relatively comparable measure of stimuli’ dissimilarity can be devised, for pairs of stimuli of different modalities, e.g., visual and spatial, respectively; and controlling for the convergence of pathways in more high-level associative areas). Due to the persistent activity-dependent changes in synaptic efficacy as well as neuronal excitability, cells that were activated by a preceding stimulus tend to be preferably recruited to process also subsequent, even dissimilar ones, occurring in temporal proximity \cite{Cai2016}, a process which, according to the present hypothesis, should be characterized by different time constants in different brain region (e.g., shorter in primary sensory areas, longer in associative ones), allowing thus to determine the relative durations of corresponding subjective moments. \par
	Addressing now more strictly what we mentioned in the previous section (as well as the question [2] above), what will determine the contribution (or lack thereof) of specific circuits, cells, or synapses (when seen from the perspective of a network they are a part of) to a conscious experience, is in our view their contribution (or lack thereof) to this very effect of ‘fusion’ of the spatio-temporal neural representations of events occurring close in time. \par
	This proposed form and role of the interaction of stimulus-related with pre-stimulus activity, relates our viewpoint to the temporo-spatial theory of consciousness (TTC), as it is a mechanism that would lead to an expansion of stimulus to points in time and space beyond the ones at which it actually occurs–a process to which TTC attributes a central role \cite{Northoff2020}. \par

\subsection{Relation of the proposed viewpoint to the extensional model of time consciousness}

	The proposed viewpoint naturally corresponds with the extensional approach to time consciousness, which posits that experienced moments are extended in time and have some duration, with such a model finding a support in empirical evidence coming from psychology and cognitive neuroscience \cite{Dainton2018,Dorato2020}. It is naturally considered in humans, and estimates that the experienced moment has a duration in the range of a few seconds, affording thus a temporal segmentation in perception and action \cite{Poppel2009,Wittmann2011,Dorato2020}. Our viewpoint would suggest to extend this approach with the evolutionary-ecological perspective and theoretical considerations about other species, and speculate that the experienced moments will tend to have such durations as to be behaviorally optimal. Based on the considerations in this section, the synaptic clocks might be seen as a natural candidate for a concrete neural basis, determining the durations of such a moments (with synaptic clock, as based on a generalized notion of synaptic tagging, or synaptic ‘eligibility trace’, directly corresponding to the ongoing phase of predictions’ evaluation and processing of ‘prediction errors’, as formulated within the predictive coding framework, see \cite{Kent2021}). In this view it can be considered that many (‘local’) moments, with varied durations, will be operating in ‘parallel’, each corresponding to an active synaptic clock. \par

\section{Consciousness as continuous change} \label{changeSection}

\subsection{The identity of experienced change and immediate memory}

	Addressing the above-discussed dichotomy between experienced change and immediate memory more strictly, it can be noted that change, by definition, is always relative to something. In the case of conscious experience, its change is relative to memory. It is not possible to conceive of any consciously experienced change without having also memory of some preceding state, or, the other way round, to conceive of memory without change. In order for some contents of experience to become a "memory" (and be recalled later) the experience needs to undergo a change. Hence, it can be stated that the constantly experienced change and, on the other hand, immediate memory (i.e., contents of experience, 'fading' into the past gradually), are in fact two different views on the same phenomenon. Without any of them, or actually without \textit{it}, the result would be the same, namely, one would be in a state that could be described as an 'eternal', i.e., not-changing, present moment. \par
	This is why the processes of neural plasticity, in particular such that can be described as being at the same time a change and a persistence of memory trace of some previous activity state, seem to be well suited to constitute content-specific neural substrates of consciousness, and this is what we shall discuss in more detail next, which should allow us to look at the above considerations from yet another perspective. \par

\subsection{The alternative views on the structure of consciousness}

	The notion of the identity of change and memory seems to be contained in the Bergson's concept of \textit{duration} \cite{Bergson1889,Bergson1896}, with \textit{duration} being the persistence of one's entire memory, accumulated over the lifetime, in light of which every new experience is interpreted (what could be labeled as 'implicit imagery'), and being constantly modified over time by integration with those new experiences. However, it appears to us that this theory, and a particular viewpoint that stems from it, cannot be understood properly or be useful without reconsidering two of its central aspects, namely (1) its treatment of the concept of memory 'storage', and, especially, (2) in what sense the term 'time' is used by it. \par
	The conventional view on the structure of consciousness (present, among others, in the IIT’s axiom of “composition”) assumes that: (1) at any given moment of time it consists of a collection of separate elements, each of which constitutes a different entity--„the percept of chair that I see on the right side of my visual field is something different than the percept of computer screen that I see in the center of my visual field”, and (2) it consists of a succession of separate collections of such elements in time, with elements at each moment of time being different entities--"the percept of chair that I am having right now is something different than the percept of that chair that I had a minute ago", which is depicted on Figure \ref{Fig3}A with such an elements represented by different points. $t$ axis on Figure \ref{Fig3}A denotes time, and $x$ axis can be thought of as representing a collection of percepts at a given moment of time, for instance reflecting different points along one dimension of a visual field. This view, in its essence, assumes that those elements are different things and can be represented as such a collection of points, with each point having different value (representing, for example, 'colors' of 'adjacent' percepts within a current visual field). This assumes that consciousness has some dimension(s) of what could be referred to as 'space' ($x$, a set of all currently experienced percepts) and a dimension of time ($t$, a succession of collections of elements, with the points taking different values, with some of them possibly becoming equal zero, e.g., no auditory input and hence no percepts of sounds). This view, however, that such elements, represented by points, are different 'things', leads inevitably to the notion of borders and empty gaps between neighboring points, that would separate different points from each other (a problem which remains even if we assume that the number of points within any interval along any of the axes is infinite). Whether it will refer to its physical substrate or to consciousness as such, it is unclear how big such a gap is, or how long it lasts, or what does it consist of, etc. Moreover, since it is by definition empty, nothing can traverse it, and thus no interaction between neighboring points and no dynamics of the conscious experience is possible. This would be in fact a situation, as described already above, of being 'stuck' in an 'eternal', non-changing present moment, without any consciously experienced change or memory. The alternative view, which is conceptualized by the notion of \textit{duration}, and which, as it seems, better describes what we experience directly, is that there are no separate points, being different 'things', with borders or empty gaps between them, but instead that consciousness is continuous (along any dimensions), namely, that it is a continuous change (Figure \ref{Fig3}B).
\begin{center}
\fbox{\includegraphics[width=\linewidth,keepaspectratio]{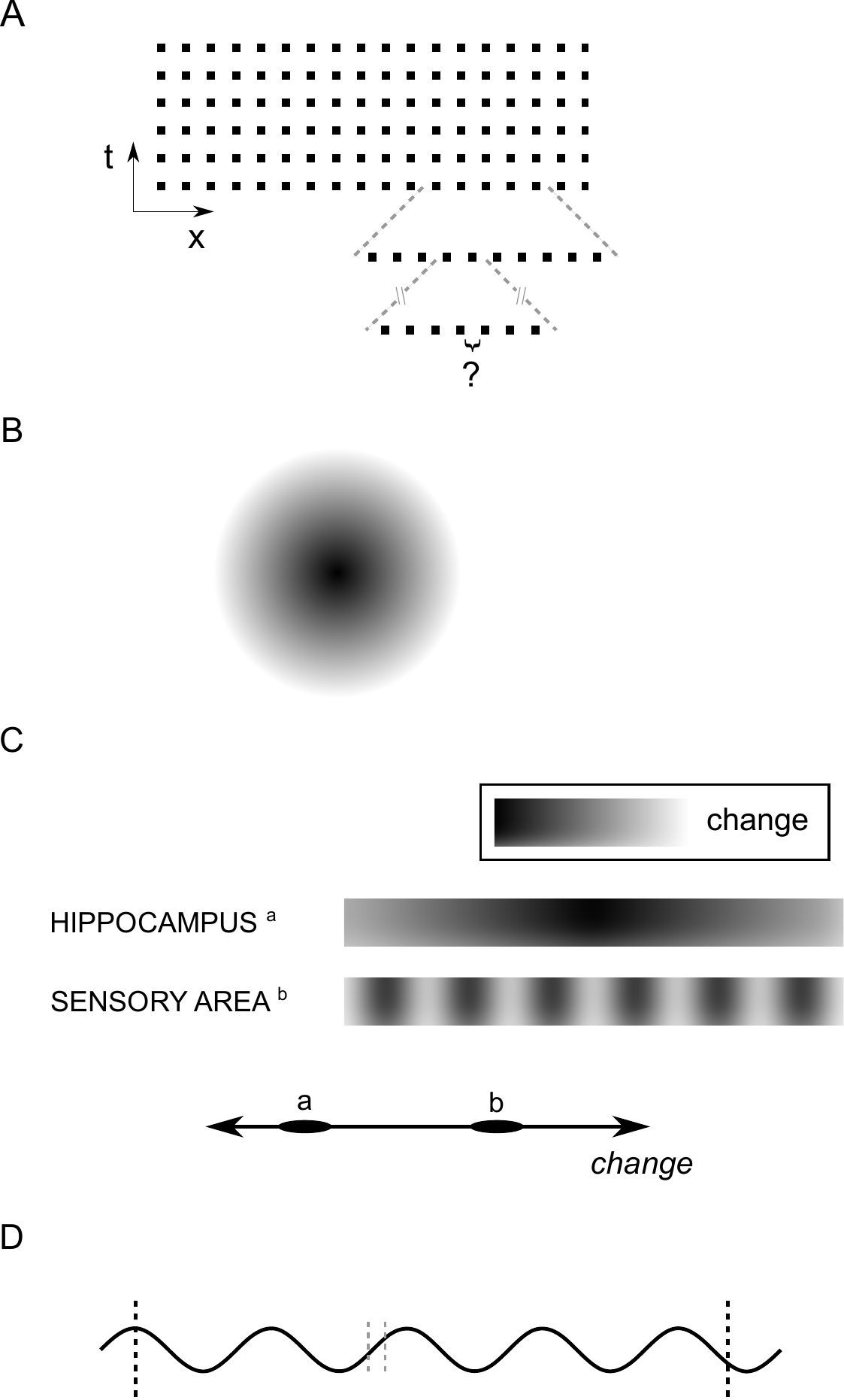}}
\captionof{figure}{Change as the dimension of consciousness. \textbf{(A--B)} Alternative views on the structure of consciousness: \textbf{(A)} static collections of elements vs. \textbf{(B)} continuous change. \textbf{(C)} Relative positions along the dimension of \textit{change}, related to different types of information processed by two example brain regions. \textbf{(D)} Perception of an element of experience (fragment in between the black dashed lines at the ends) vs. perception of the process of its generation (fragment in between the gray dashed lines in the middle). \label{Fig3}}
\end{center} \par
	However, if this is indeed the case and a useful theory is to be built upon this notion, then it seems that, first of all, the use of term 'time', and 'space' as well, is inadequate and, moreover, might be somewhat misleading, as is the case with the \textit{duration} theory. What appears to us as a constructive approach is therefore to consider \textit{change} itself as the only dimension of consciousness. \par

\subsection{The dimension of change}

	However, once this is recognized, the conclusion is that, in fact, there are no separate elements experienced at a given moment of time, nor is there a linear flow of conscious experience along an axis of time, as it is commonly thought of, with separate past, present and future time moments, but instead that consciousness is always, in a sense, in the same moment of time and that all experiences are in fact one whole. This situation is not identical to the one considered above, namely of an 'eternal' non-changing moment, nor is it a form of 'presentism', according to which time does flow but only present moments are actually real \cite{Dainton2018}, as this is, strictly speaking, not a moment of 'time', or a 'spatially' separable fragment of experience, but a constant \textit{change}. In other words, consciousness has no discrete 'spatial' or 'temporal' dimensions, nor is there a distinguished dimension of 'time', along which something would proceed. The difference between the conventional notion of dimension of time and that of \textit{change} is that a single static point located anywhere along the dimension of \textit{change} represents, in itself, a change, in the sense as outlined above, namely, a specific continuous change relative to immediate memory. Whereas in the case of dimension of time, such a static point would always represent \textit{no change}, in any quantity measured. As behaviors of any system are typically considered in terms of changes of some quantity with \textit{time} (i.e., along the dimension of time), and the goal then is to account for its dynamics, the reasoning outlined here may thus suggest that this kind of analysis is not possible. And it appears to us that indeed this might be the case, and that ultimately what can be only studied about consciousness is its 'kinematics', i.e., constructing statistical descriptions of how it behaves, or, in other words, what observable effects it produces (à la Best System Analysis approach to the laws of nature \cite{Hoefer2016}). However, since we have a direct access to our consciousness from the 'inside', meaning that we actually experience it, and are able to reflect on it, we posit that what might be an especially promising step forward is the fact of the variability of rate of subjective time passage that we perceive, which in combination with indirect knowledge about that rate in other species \cite{Healy2013}, might be informative, first of all, as to what does it mean to move along the axis of \textit{change}, as depicted on Figure \ref{Fig3}C. \par
	The rationale for speaking about consciousness as having a dimension of \textit{change}, rather than being simply a discrete point of \textit{change}, is that since it is continuous, it seems that it cannot be confined to a static, i.e., discrete, point of a constant 'rate', but instead it will at all times 'accelerate' or 'decelerate', as is suggested by the variability of subjectively perceived time flow, which can be represented as motion along the axis. \par
	The model from Figure \ref{Fig2} could be now modified in a manner as depicted in Figure \ref{Fig3}C, with more 'rapid' changes occurring according to the sensory area, as compared to the hippocampus' judgment, according to which much less change has occurred within the same interval of objective time (and we are using here the time-related grammatical forms--"occurring", "has occurred"--only because of the lack of a more suitable language). However, the varied 'rate' of \textit{change}, that is what is represented as the one-dimensional axis of \textit{change}, is not quantitative, as it is by definition a continuum. It can be considered as such only in light of a retrospective reflection on a given experience, as a way of abstract description of the actual experience (assessing "how many changes occurred"). Instead, from the subjective perspective of actual experience, \textit{change} does not have any particular rate, it just is the way it is, and its 'rate' is something we can infer only retrospectively by comparing it with what we then consider to be some different fragments of experience. The dimension of change is thus to be treated rather as an abstract tool, offered by an act of introspection (ability to manipulate what we consider to be separate pieces of experience), whereas an actual experience simply changes in some specific way inherent to it. In other words, quantitative descriptions, using the language of mathematics, based on the notion of mathematical point, are incompatible with the essentially qualitative reality of conscious experience. \par
	The act of abstraction and comparing past experiences appears to rely on memory of those experiences, that must be somehow stored and preserved over time since the events originally occurred, which concept we shall consider next, analyzing whether and how it can be reconciled with the notion of continuous \textit{change}. \par

\subsection{Are memories stored?}

	The notion of continuous change seems to contradict our conviction that there are separate elements in every conscious experience, constituting different entities, and that time flows linearly, i.e., from past to future, with past events, memory of which has been retained, having logic continuation and consequences in present events. The concept of memory 'storage', on which this view is based, assumes that events that were experienced in the past, i.e., in some previous moment of time, can be somehow 'saved’, in a form of memory trace, instead of ceasing to exist, and then retrieved from that store in some present moment of time, namely, that what is considered a memory recall and an event that is recalled can all be marked on a line, or an array, like the one on Figure \ref{Fig3}A, with past moments of time, when the original memorized event took place, being something different than the present moment, when it is recalled. However, what does it mean exactly that memory is 'stored', and what evidence there is to support such notion? \par
	Studies on mechanisms underlying memory, with memory being assessed mostly in behavioral paradigms in animals, suggest that memories are stored in the brain in a form of neural memory traces, termed "engrams" \cite{Tonegawa2015}, being specific patterns of activity of neuronal ensembles requiring plastic synaptic changes to be encoded and then replayed, that constitute representations of some particular events from the past. However, it can be noted that every instance of what is thought of as recall of a memory, always occurs in a context different than the one in which it was supposedly formed, resulting in an overall experience being different than (i.e., altered relative to) the original one (even for the very reason that this is an event of "memory recall", accompanied by an awareness of this act of remembering), and it always takes place in a present moment of time. Moreover, the flow of conscious experience seems to be continuous, without any clear-cut borders between past and present moments, that would separate events that were memorized from subsequent remembering of those events. Neural plasticity, to which the formation of memory is attributed, seems to result primarily not in the formation of representations of particular events, but rather in adaptations to ever-changing situational demands \cite{Hardt2020}. Technically, any activity-dependent act of synaptic plasticity makes it less, not more, likely--or actually, considering an entire system, impossible--that an identical overall pattern of network activity will be replayed in the future (Figure \ref{Fig4}). Thus, we do not have any direct evidence to support the notion that anything is 'stored', and can be then recollected, or that there is any separate 'past', as commonly thought of assuming a linear progression of time moments. We have a direct evidence only that our conscious experience changes constantly. Especially in light of the fact that in the brain there seems to be no separation of sites of information processing from the sites of memory storage. Any memory-related processes (be it encoding, consolidation, maintenance, retrieval, reconsolidation, forgetting, etc) are all constituted by some information processing in the brain that takes place, and affects the functioning of organism, in the ‘present’. There is thus no evident need to assume that what we experience as "memory" is something that was stored (anywhere, be it brain or mind) at some point in the past.\begin{center}
\fbox{\includegraphics[width=\linewidth,keepaspectratio]{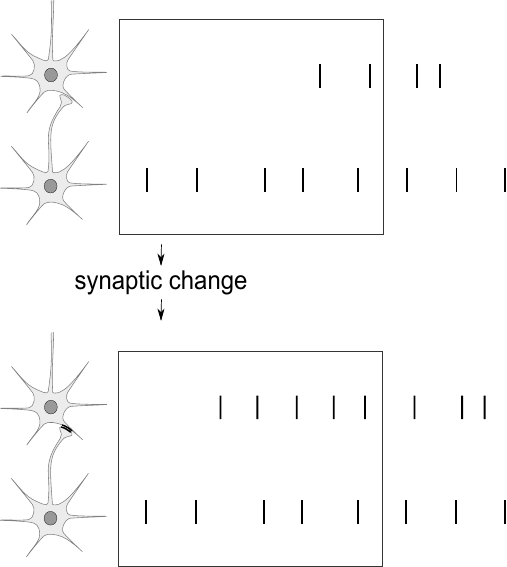}}
\captionof{figure}{Illustration of the effect of an activity-dependent synaptic modification on the network activity patterns. Once the weight of the synaptic connection between a pair of cells is changed, the replay of an overall pattern of network activity identical to the one that was related to an original experience becomes not possible. Now, as soon as the input cell starts firing, its activity 'attracts' the activity of the output cell, resulting in a different overall pattern of network activity. \label{Fig4}}
\end{center} \par
	It looks like the conventional view on time is based on the static notion of memory (‘static’, in the sense that even though in this view memory traces can be altered or erased, they are nonetheless assumed to represent particular events, referring to particular time points in the past), which in turn is based on the notion of a linear flow of time, both of which are, as we have attempted to show, not adequate as a descriptions of consciousness. It appears to be a case of circular reasoning, which is resolved by the adoption of the dimension of \textit{change}. Accepting that no 'storage' (in the conventional sense of this term) of memory ever occurs eliminates the problem (and that is not to say that it is not practical to treat and study the cognitive and behavioral processes of memory 'access' as if memories were indeed, in some sense and in certain approximation, stored \cite{Howard2015}). \par
	But then, what makes the contents of my experience appear to me like they represent an event from the past being remembered in some present situation?--a question that reduces to a more basic one: what constitutes the 'meaning' of any one conscious experience (allowing to distinguish for example a remembering of an event from the perception of an actual event)? \par

\subsection{Seeing or remembering: On the meaning in the contents of a conscious experience}

	One could expect that such meaning should be a result of the intrinsic structure of a given conscious experience, in which, if it is taken in its entirety, there is contained information about different elements of the experience and about relations between them, i.e., how they are arranged. For instance, that there was a past when a given event took place, and that it is remembered now in the present while different percepts are also being perceived forming a coherent image of the current surroundings, and so on, with the contents of such experience acting thus like a static 'time capsule', by analogy to fossils, which can be seen as 'memories' indicating that there was some past in which they were created \cite{Barbour1999}. The same reasoning would extend to visualizing the future, which is always done using some elements from memory, and is done in the present, constituting an actual experience. Such a meaning would allow to compare also durations of different past intervals, i.e., assessing the number of changes that occurred, and apply as well to any other experience constituting hence a coherent whole (with phenomenal consciousness and cognitive access to its contents being in this view synonymous concepts, with every instance of "cognitive access", or a failure of one, constituting actually another experience \cite{Phillips2018}). However, in contrast to fossils as such, whose 'meaning' needs to be derived by an external observer, our experiences seem to be self-interpreted, suggesting that such a static images are not sufficient. \par
	We argue that what makes this kind of meaning, being intrinsic in the contents constituting an experience, possible, is first of all that they are not static, but rather continuously \textit{change}. In the case of a static image we come back essentially to the array from Figure \ref{Fig3}A, that is a collection of points, in which situation it is unclear whether one's experience is confined to only one particular point or to all of them at the same time, which anyway does not lead to the construction of a coherent whole, regardless of how structured such an image, constituted by a collection of separate elements, would look from the outside, for an external observer. And if any meaning cannot be derived in case of any of such collections individually, it seems to be not possible also to derive it from a set of collections, each without any meaning (as such an attempt would be like looking along yet another dimension of the array). We posit that what enables the construction of experiences with this kind of meaning, is a dynamic, immediate process of collecting information within 'frames' of continuous \textit{change}, which could be described as sampling, and integration of this information into a meaningful whole--a process of continuous learning, which we shall consider below. \par
	In the view outlined, what we consider an explicit recall of a memory, has the same function as any percept. It serves to guide current behavior. It is thus, as a consequence, reasonable to assume that there is no fundamental difference between structured memories and percepts. This point of view, and analyzing how these two phenomena are related and what they have in common, in terms of their 'temporal' aspect, might be particularly informative for and will lead us eventually to addressing an issue: what constitutes the difference between subject and object, after we accept that the dimension of consciousness is \textit{change}? \par

\subsection{How to deal with panpsychism}

	It is argued by some that the fact, or actually--feeling, of the flow of time might be not only the most basic aspect that our consciousness shares with mental lives of individuals from other species, as discussed above, but also be the only aspect that it shares, in some way, with the physical reality, as formulated in a view assuming that time is real and plays a central role (as opposed to the concept of a block universe in which nothing \textit{really} changes) (see, for example, \cite{Smolin2015,Dainton2018}). This perhaps could be understood within some form of the panpsychism view, according to which something resembling what we know as "consciousness" constitutes the intrinsic nature of matter \cite{Goff2017,Lamme2018}. We expect that if this is indeed the case, then considerations of the physical reality would benefit from adopting a reasoning as the one outlined above, as purely logical argument, and putting \textit{change} as this reality's only dimension. \par
	Taking this particular perspective, we shall then adopt a view, according to which the difference between a perceiving subject (i.e., one's conscious experience) and perceived object (i.e., objective reality) lies not in spatial, but rather in the temporal domain \cite{Bergson1896}. However, as we have resigned from using the term 'time', or 'temporal', arguing why it is inadequate, we posit that this should be defined rather in terms of \textit{change}. \par

\subsection{Direct perception}

	 The view on the difference between subject and object, when defined in 'temporal' terms, posits essentially that a subject is wherever the 'pure' (i.e., autobiographical/episodic) memory of all of the subject's past experiences are stored, whereas objects are located always in the present, in the currently unfolding moment \cite{Bergson1896}. We diverge from this view in assuming that since memory is always manifested in the present, there is no evidence or need to assume its 'storage', or to assume a notion of static past, that would need to be taken into considerations. \par
	Our approach, however, still suggests taking as a starting point the theory of 'direct' perception \cite{Bergson1896,Gibson1972}, which posits that all percepts are images, that essentially are some selected fragments extracted, in a sense, from an overall continuous stream of information that the physical reality is--an ability which evolves, and is developed in and, to a lesser degree, learned by individuals. \par

\subsection{Sampling and integration of information}

	The theory of ‘direct’ (‘ecological’) perception \cite{Gibson1972}, assumes that the perception of a specific environment's structure arises due to sampling, in which visual information is sampled over time, in a process dependent on a moving focus of attention, and integrated in a certain way, in a loop which involves motor reactions allowing to sample different perspectives of the environment, leading to the extraction of features that are invariant, in order to produce a coherent image guiding possible actions in that environment. It seems that the maximum rate of such a sampling will be proportional to the value of CFF in a given species. Whereas visual system as a whole samples continuously, the resolution of its basic elements will be limited by the CFF values. \par
	In terms of neural substrate of perception, this theory suggests that the content of what is being perceived at a given moment, due to perception being a prolonged process, will be dependent on an activity not only in the retina and primary visual areas, but rather it will involve also activity of certain motor circuits, that altogether constitute a perceptual system, with the engagement of specific neural substrates being modulated by the attentional processes, i.e., attention focused on different portions or aspects of the environment. Taking into consideration the fact that even minor parameters of brain activity can affect the functioning of the brain as a whole \cite{Tononi2016,Li2009}, it can be expected that what is actually being perceived, in details, will be determined by collective activity patterns that may involve, to varied degrees, networks of the entire brain. \par
	We posit that other types of information, processed by different systems, are sampled and integrated in a similar manner, within temporal windows of duration specific to them, as represented by the time units of synaptic clocks, which allows a given system to form a meaningful 'image'. For example, resorting to Figure \ref{Fig2}, with the hippocampus receiving inputs from sensory areas, it will construct a single image integrating many images constructed at a higher rate (e.g, ~60/sec) by the afferent areas. But also in the opposite direction, if a sensory area receives input from the hippocampus (possibly an indirect one), it will construct its consecutive images using information about a single image being constructed by the hippocampus. Consequently, the contents of conscious experience related to the activity of a region will be determined by the informational content integrated over the temporal window of duration specific to that region, through a process that will include also immediate feedback information about the effects that the activity in this temporal window is having on the organism. Contents of a conscious experience as a whole will arise from the sampling and integration of information by all subsystems, which perhaps could be studied using an approach like that of IIT \cite{Tononi2004}. \par
	This viewpoint naturally entails the embodied and proactive theories of mind, according to which conscious experience is not a result of representational processes in the head, but rather is determined by sensorimotor activity of the whole body which proactively navigates the world and is dynamically coupled with the environment, which might determine the dynamics of subjective experience of time flow \cite{Montemayor2021}. \par

\subsection{Two sides of conscious experience}

	The processes of generation of percepts, including such being the elements of a memory recall, as well as of any other contents of experience, have already been either evolved, i.e., they have been evolutionarily 'mastered', or learned, and now they are unconscious and seem to be immediate, not requiring any effort. Namely, a given content of experience simply appears in consciousness at some point, e.g., as in the comprehension of speech where individual words simply appear in consciousness. We posit that all such contents of experience, including memories, are, similarly to percepts, fragments that are 'extracted' from the physical reality as a whole, and stabilized in an apparently unchanged form for some period of time. They thus constitute specific 'paths' paved in \textit{change}, out of all possible paths that could be there at the moment. \par
	Whereas, in contrast to the above, what is conscious is a process of generation of one's memory as a whole, that is, a process of generation of behavioral adaptations. This is what we experience directly, as continuous learning. The actual experience, that appears to us as unfolding in time, is like one were 'inside' of the process of generation of some specific contents of experience (e.g., a percept). ‘Zooming in’ that process, one can see it unfolding, i.e., changing in a specific way, seeing thus its fine structure (which is like being confined to a specific, exact point in time), but is then unable to perceive its global structure and thus will not see that percept as a whole (by analogy to the uncertainty principle, as known from time-frequency analysis of time series; as illustrated on Figure \ref{Fig3}D). \par
	In sum, there are two sides from which any one conscious experience can be considered, corresponding to immediate memory and change, respectively (as described above): when we look at it in a 'retrospective', abstract manner, at any stage, we see some specific contents of experience already formed, persisting in consciousness for some period of 'time'. Processing of different types of information, related to different contents of experience, moves us then along the axis of change (Figure \ref{Fig3}C), proportionally to the corresponding values of ‘CFF’, affecting differently our assessment of the duration of that experience and thus speed of time flow. However, looking at it from the other side, when the experience is actually unfolding, without our reflecting on it, then each such window, that would be proportional to 1/CFF, lasts simply as long as it does and its duration cannot be measured using any arbitrary units, and it constitutes a process of continuous learning. \par	

\subsection{How the brain learns to be conscious}

	In the sense as outlined above, a neural memory trace, encoded in a network in a set of synaptic connections, can be seen as a structure that enables an individual to extract and 'perceive' a specific information in the physical world. Therefore, an analogy can be drawn between the process of generation of memory traces (engrams) and an evolutionary process of generation of structures enabling the organisms to perceive a specific type of information in their environment, i.e., to have a specific type of percepts (e.g., of a red light), with inadequate structures, and traces, being eliminated (one might object that an episodic memory contemplated in daytime is not something that can be considered equal to a percept used pragmatically in some specific situation to navigate in the environment. However, we posit that, if considered in light of \textit{change}, they both equally serve to accomplish pragmatic goals in respective specific circumstances, e.g., when using some elements from memory in order to solve a mathematical problem. The point is that individuals are evolutionarily adapted, and learn, to daydream only in situations in which this type of behavior is desirable, and not in situations of an immediate danger and need for a directed action). This analogy can be defined in particular on the level of plasticity of single synapses, with emotions acting as selective pressures, eliminating neural memory traces that are not desirable for the brain. \par
	Conscious experience is always associated with emotional states. Such state can be a fear, or at least a feeling of discomfort, or a pleasure, or at least a feeling of comfort, but it is never neutral. As discussed already above, emotions modulate learning, in particular through the activities of neuromodulatory systems, modulating neural plasticity on the level of synapses, thus shaping the structure of neural networks. Some circuits may be affected to a lesser degree, when related behaviors are automatic and usually do not need to be adjusted (e.g., primary sensory and motor pathways), and others are more affected, however all seem to be susceptible to such a modulation \cite{Molina-Luna2009,Tropea2009}. It is thus like the brain was organizing itself, through the neuromodulatory effects related with emotions, or actually a sort of micro-emotions \cite{Dennett2018}, restricting how it \textit{changes}, by changing its memory as a whole, and thus selecting what information exactly it is capable to 'perceive', i.e., what contents of experience it can have. Subjective emotional states, indicating whether organism is in a desirable or a non-desirable overall state, are present regardless of what are the contents of a given experience. Emotions seem to permeate every experience. Whereas, in contrast, it is possible to think (although it does not seem possible to imagine, or feel) that objective reality \textit{changes} only in some neutral, 'random' way. We posit therefore that the difference between subject, i.e., our subjective experience, and object, nature of which might itself resemble our consciousness, should be sought rather in the domain of (micro-)emotions, in how they organize the components of a physical system so that it changes in a desirable way. That is, such a way, that it is able to perceive those pieces of information, e.g., as explicit memories or through more implicit imagery, that give it most adaptive advantage. In this sense it can be said literally that the brain learns (and, also, teaches itself) to be conscious \cite{Cleeremans2011}, and we posit that the synaptic clock represents an elementary process through which this occurs. Namely, we posit that synaptic clocks determine 'temporal' windows (which, however, should be understood in light of the dimension of \textit{change}) in which the following occurs: (1) information is sampled and integrated, with the type of information depending on the region, which leads to the extraction of invariant features within the sample, (2) the sampling and integration in this prolonged window depends on the activity of the organism, its behavior, and feedbacks it receives from the environment, (3) the result of functioning of all the clocks, as well as of each clock in particular, is a meaningful 'image', constituting some contents of experience (e.g., of remembering an event), (4) the effect that those contents are having on the organism is continuously evaluated by dedicated systems, which leads to (micro-)emotions, reflecting the brain's 'opinion' on its present state, (5) emotions act through neuromodulation, and PRPs, continuously affecting the synaptic (or, in general, neural) changes, (6) the modulation of synaptic changes alters the routes of signals' flow through the network, and thus content of information sampled. \par

\section{Discussion}

\subsection{Few more arguments for the synaptic clock mechanism}

	In support of the existence of a general mechanism of synaptic clock, based on a generalized notion of synaptic tagging, in terms of its possible adaptive values, also the following arguments could be put forward:
\begin{itemize}
	\item Modeling studies on artificial neural networks show that such networks can develop complex computational functionalities \cite{Hoerzer2012}, or learn to guide behavior in simulated robotic applications \cite{Tsai2018}, when they are trained using learning rules based on reward-modulated Hebbian-like plasticity with a single reward signal. In such paradigms the activity of a network in previous steps of time makes the synaptic connections involved in that activity eligible for subsequent reinforcement, with what can be in general considered synaptic tags. One could expect that developing a distribution of synaptic traces whose durations of persistence are varied and adjusted specifically to enable the action of such signals on particular synapses, being first of all not too short but also not too long, would be even more effective for specific learning purposes and thus beneficial for organisms (and perhaps even more so with suitably shaped trace-decay functions, as envisioned for neural eligibility traces by Klopf, see \cite{Sutton2018});
	\item Synaptic learning and memory is estimated to be relatively cheap energetically, in terms of metabolic costs of sustaining a synaptic memory trace \cite{Karbowski2019}. One could therefore expect that the durations of persistence of such synaptic memory traces, in particular of their initial stage corresponding to synaptic tagging, can be shaped evolutionarily in a flexible manner, where longer-lasting, behaviorally not immediately beneficial traces can be 'tested out', not being under a strong requirement of limiting the energy expenditure;
	\item Activity-dependent synaptic plasticity and LTP in particular might serve different functions in different brain systems, that is, encode different types of memory and thus give different results in terms of specific circuits' functioning and cognitive processing \cite{Takeuchi2014}. Correspondingly, through one general mechanism of "synaptic clock" various effects on the networks' activities and associated cognition could be realized. If time is a universal ecological dimension, the existence of varied distributions of instances of "synaptic clock", that are shaped by how organisms interact with their environments, especially by rates of those interactions, being adjusted to their specific needs, would suffice different organisms and different brain systems, in this sense making it a parsimonious explanation;
	\item At the moment of a synaptic event corresponding to an experience that should be memorized, the availability of various PRPs, potentially needed to actually implement the synaptic change, might be varied, and they can be at different stages of their production process \cite{Costa-Mattioli2009,Graber2013}. It is conceivable that in different synapses, with synaptic traces of different durations, there might be a varied dependence of plasticity on the PRPs at different stages of the production process. Namely, activity-dependent synaptic plasticity in synapses with gradually longer time units of synaptic traces could be less dependent on: only already-synthesized proteins immediately available at the synaptic site (produced due to some preceding events)\textrightarrow mRNAs whose translation has been paused at the elongation stage and can be reactivated on demand\textrightarrow mRNAs before translation initiation\textrightarrow \textit{de novo} transcription. Varied time units of synaptic traces would thus lead to varied dependence of activity-dependent synaptic plasticity on the history of a given synapse' and cell's activity. Although it is rather unclear whether this would in itself constitute a desirable feature.
\end{itemize} \par

\subsection{Synaptic clock as a neural substrate of consciousness vs. possibly no consciousness in certain states or systems}
	
	How to reconcile the proposed mechanism of synaptic clock, or some related mechanism based on neural plasticity, as a substrate of consciousness, with cases of certain systems or situations in which, possibly, there is no consciousness associated with them? For instance, in case of (1) cerebellum, which, as it seems, does not give rise to any evident conscious experiences, (2) dreamless sleep, or (3) lower animals, e.g., marine mollusks (like \textit{Aplysia Californica}), which may or may not be conscious--despite substantial amounts of neural activity and plastic synaptic changes in all of them \cite{Koch2016,Takeuchi2014,Martin2002}? \par
	Our main argument is that we aimed to find an account for certain properties of consciousness, and what is proposed is that whenever consciousness is possible to occur the persistence of its contents will correlate with some synaptic traces (or some analogous processes; and possibly with only a subset of them). Not that synaptic plasticity is in itself sufficient for any conscious experience to occur. Importantly, the present proposal should be understood from the above-described broader perspective of networks, activity of which the synaptic plasticity shapes. For consciousness to be possible, what is perhaps necessary are other mechanisms that constitute predispositions for, and prerequisites of, the actual proper neural correlates of consciousness \cite{Northoff2020}. \par
	Also, some additional lines of reasoning could be proposed for two of the above cases. Namely, for: \\
\indent (1) cerebellum--this structure is organized into largely non-overlapping functional modules, receiving mapped inputs from the environment, with little interaction possible between separate modules \cite{Tononi2004}. This property, combined with the fact that it may have the putative synaptic clocks with the time units so short that almost non-persistent (since it can be considered a part of low-level sensory/motor circuitry), will prevent this structure from contributing (strongly) to the above-described ‘fusion’ of spatio-temporal representations of non-co-temporal inputs. As a result, its impact on conscious experience will be not 'noticeable'; \\
\indent (2) sleep--from the perspective of one's conscious experience the dreamless sleep can be viewed as a period in which the perceived time flow speeds up so that this period seems to last ‘infinitely’ short (when assessed in retrospection). It is thus like the moment just after awakening was a continuation of the one just before falling asleep or, alternatively, the one just before the last dream experienced during the night ended (and because of that one cannot be sure when exactly the dream occurred, that is, whether it was just before the awakening or, maybe, just after falling asleep). Hence the supposed lack of consciousness during the dreamless sleep could support the view that when no low-level sensory data is being processed in an integrated manner, and no voluntary movements executed, the remaining type of information processing speeds up the time flow so substantially. Addressing this issue of sleep more strictly, this ‘temporal’ situation of dreamless sleep seems to be analogous to a ‘spatial’ one, of the blind spot in our visual field, existence of which we are not directly aware of, unless we are informed about the properties of a putative underlying physical system--then we are ‘aware’ of it, but still not directly. In other words, a lack of (supposed) information (as in these two examples) is not the same as would be an information about a lack \cite{Dennett1992,Dainton2018}. \par

\subsection{Continuous change and the metaphysics of time}

	We propose that, due to reasons as outlined above, the continuous \textit{change}, when seen as the dimension of physical reality, would constitute a desirable alternative (see \cite{Wittmann2021,Montemayor2021}) to the current models of the large-scale structure and composition of time itself (i.e., the time of the universe as a whole), with those models assuming either no real passage of time (as in the block universe model) or some forms of objective cosmic passage (see \cite{Dainton2018}). \par
	What we consider 'space' and 'time' could actually be abstractions derived from the two opposite extrema of the \textit{change} axis, with the notion of 'space' derived from a small amount of change (an almost constant configuration of objects, not changing over time) and the notion of 'time' derived from rapid changes (what can be imagined as if everything in our conscious experience was changing constantly in a random manner, without any recurring elements or regularities, then we could not conceive of or have any conception of space, just of 'pure' time). \par
	Meditation practitioners report experiences described as “timelessness” (commonly occurring jointly with an experience of “spacelessness”) \cite{Berkovich-Ohana2013}. The term “timelessness”, however, is used to refer to specific experiences of an altered sense of time, that is experiences of being “outside of time”, as opposed to being “now”, inside a present situation. This, in our view, does not entail any notion of the flow of subjective experience actually stopping (or ‘pausing’, and then ‘restarting’). \par

\section{Conclusion}

	To sum up, in the present paper we considered the temporal aspect of consciousness from an evolutionary-ecological perspective, and proposed (1) a synthesis of certain elements of different current theories, (2) that synaptic clock might constitute a content-specific neural substrate of consciousness, (3) that many moments of subjective time, with different durations, might be considered to operate in ‘parallel’, (4) continuous change as the dimension of consciousness, accounting thus for the subjective experience of time flow. \par

\section*{Acknowledgments}

I thank Marc Wittmann, and Giorgio Marchetti, for helpful comments and suggestions on the manuscript.

\end{multicols}
\end{document}